\documentclass[preprint,superscriptaddress,showpacs,preprintnumbers,amsmath,amssymb,pre]{revtex4}
\usepackage{epsfig}
\usepackage{latexsym}
\usepackage{dcolumn}
\usepackage{subfigure}
\usepackage{bm}
\begin{document}
\date{\today}
\title{Fractal Heterogeneous Media}
\author{Christian T\"{u}rk}\email{christian.turk@polito.it}
\affiliation{Physics Department and CNISM, Politecnico di Torino,\\ Corso Duca degli
Abruzzi 24, I-10129 Torino, Italy}

\author{Anna Carbone}\email{anna.carbone@polito.it}
 \affiliation{Physics Department and CNISM, Politecnico di Torino,\\ Corso Duca degli
Abruzzi 24, I-10129 Torino, Italy}

\author{Bernardino M. Chiaia}\email{bernardino.chiaia@polito.it}
\affiliation{Department of Structural and Geotechnical Engineering, Politecnico di Torino,\\  Corso Duca degli Abruzzi 24, I-10129 Torino, Italy}

\begin{abstract}
 A method is proposed for generating compact fractal disordered media, by generalizing the random midpoint displacement algorithm.  The obtained structures are  \emph{invasive stochastic fractals}, with the Hurst exponent varying as a continuous parameter, as opposed to \emph{lacunar deterministic fractals}, such as the Menger sponge.  By employing the Detrending Moving Average algorithm [Phys.~Rev.~E~\textbf{76}, 056703 (2007)], the Hurst exponent of the generated structure can be subsequently checked.
The fractality of such a structure is referred to a property defined over a three dimensional topology rather than to the topology itself.  Consequently, in this framework, the Hurst exponent should be intended as an estimator of \emph{compactness}  rather than of \emph{roughness}. Applications can be envisaged for simulating and quantifying complex systems characterized by self-similar heterogeneity across space. For example, exploitation areas range from the design and control of multifunctional self-assembled artificial nano and micro structures, to the analysis and modelling of complex pattern formation in  biology,  environmental sciences, geomorphological sciences, etc.
\end{abstract}
\maketitle
\section{Introduction}
Real world materials, like stable and metastable liquids, glasses, defective crystals, and structures that evolve from non equilibrium processes,  always exhibit a certain degree of disorder.
On the other hand, the design of artificial heterogeneous structures and the accurate control of their structural, chemical or orientational disorder is an area of intensive investigation for fundamental and technological interest \cite{Lookman,Gommes,Nicodemi,Fullwood,Mladek,Corte,Kamien,Briscoe,Jerkins,Williams,Hilfer,Ketcham,Knackstedt}. Recent developments in nanophotonics  have  shown that it is possible  to make use of the intrinsic disorder in natural and photonic materials to create new optical structures and functionalities
\cite{Wiersma}. Fractal bodies, such as Menger sponges \cite{nota}, have demonstrated the ability to perform novel functions as  localize electromagnetic and acoustic waves  \cite{Takeda,Sakoda,Hou} or enhance super-liquid repellency or dewettability \cite{Mayama,Minami}.
The reconstruction of heterogeneous media, from the knowledge of the correlation function, is a challenging  inverse problem. Any reconstruction  procedure requires to be effectively controlled in order to enable the prediction, design and implementation of structures exhibiting the desired electromagnetic, transport or biological functions \cite{Debye,Yeong,Jiao,Scardicchio,Arns,Quintanilla,Aste,Wang}.
\par
The disorder degree of a medium can be quantified in terms of the two-point correlation function $C(\mathbf{r_1},\mathbf{r_2})=<f(\mathbf{r_1})f(\mathbf{r_2})>$ of a relevant quantity $f(r)$ (e.g. dielectric function, porosity, density), where $\mathbf{r_1},\mathbf{r_2}$ are two arbitrary points in the system. For statistically isotropic media, $C(\mathbf{r_1},\mathbf{r_2})$ depends only on the distance $\lambda=\|\mathbf{r_2}-\mathbf{r_1}\|$ between two points, thus is written as $C(\lambda)$.  For fully uncorrelated systems, like the ideal gas, the correlation function is a simple exponential, $C(\lambda)\propto \exp(-\lambda/a)$, while for fully ordered media, like the perfect lattice, the correlation function $C(\lambda)$ is a constant. Generally, heterogeneous materials exhibit forms of correlation which are intermediate between those of the ideal gas and the perfect lattice.
\par
Fractional Brownian functions  are characterized by a correlation function  depending as a power-law on  $\lambda$ \cite{Mandelbrot,Falconer}. Such a correlation  may reasonably account for the intermediate behavior,  between the fully uncorrelated exponential and the fully correlated constant decay,  exhibited by real disordered media.
The power-law correlation of fractional Brownian functions
  can be expressed by the power-law decay of the variance:
\begin{equation}
\left \langle [f_H({{r}}+ {\lambda})-f_H({r})]^2
\right \rangle \,\,=\sigma_o^2 \|{\lambda}\|^{2H} \hspace*{10 pt}, \label{variance}
\end{equation}
\noindent with $f_H({r}):\mathbb{R}^d\rightarrow \mathbb{R}$, with $ {r}=(x_1, x_2,...,x_d)$, ${\lambda}=
(\lambda_1,\lambda_2,...,\lambda_d)$ and  $ \|{\lambda}\|=\sqrt{\lambda_1^2+\lambda_2^2+...+\lambda_d^2}\,$.
$H$ is the  Hurst exponent, which is related to the fractal dimension through the relation
$D=d+1-H$, with $d$ the embedded Euclidean dimension.
$H$ ranges from $0$ to $1$, taking the values $H=0.5$,
$H>0.5$ and $H<0.5$ respectively for uncorrelated, correlated and
anticorrelated Brownian functions.
Concepts as scaling, criticality and fractality have been proven useful
to model dynamic processes \cite{Aguirre}; stress induced morphological transformation
 \cite{Blair};  isotropic and anisotropic fracture surfaces
\cite{Ponson,Hansen,Bouchbinder,Schmittbuhl,Santucci}; static
friction between materials dominated by hard core interactions
\cite{Sokoloff}; elastic and contact properties \cite{Li,Carpinteri1,Persson,Carpinteri2,Pradhan,Mo}; diffusion and transport
 in porous and composite materials  \cite{Levitz,Malek,Oskoee,Filoche}; mass
fractal features in wet/dried gels, liposome and colloids \cite{Vollet,Roldan,Manley};
physiological organs (e.g. lung) \cite{Suki}; polarizabilities \cite{Nakano}, hydrophobicity of
surfaces with hierarchic structure undergoing natural selection
mechanism \cite{Yang} and solubility of nanoparticles
\cite{Mihranyan}. Several quantification methods have been proposed to accomplish accurate and fast estimates of power-law correlations at different scales \cite{Rangarajan,Davies,Alvarez,Gu,Kestener,Carbone1,Carbone2,Carbone3,Arianos}.
\bigskip
\par
In the present work:

\begin{enumerate}
\item A compact  invasive stochastic  fractal structure is obtained  by generalizing the random midpoint displacement (RMD) algorithm to high-dimension (Section II). In topological dimension $d=3$, a fractal cube with size $N_1 \times N_2 \times N_3$ is obtained, whose relevant property (e.g. density, dielectric function, porosity) is described as a three-dimensional fractional Brownian field $f_H (r)$. As opposed to deterministic lacunar fractals, such as the Menger sponge whose fractal dimension is equal to  $2.7268$, the obtained medium  is a stochastic invasive fractal. By varying the Hurst exponent, which is given as input of the algorithm, between $0$ and $1$, the fractal dimension $D$ is continuously varied  between $D=3$ and $D=4$. It is worthy of remark that in this case the fractal dimension  $D$ refers to the \emph{compactness}  rather than to \emph{roughness} as in the case of  fractal surfaces and interfaces. Such a structure can be used for describing complex materials with correlation function exhibiting a power-law dependence over distance  and arbitrary degree of heterogeneity expressed by $H$.

\item  The degree of disorder of the  medium is quantified in terms of the Hurst exponent, whose estimate is provided by the Detrended Moving Average  (DMA)  algorithm \cite{Carbone1} (Section III). The algorithm calculates  the generalized variance $\sigma_{DMA}(s)$ of the
fractional Brownian fields $f_H (r)$ around the three-dimensional moving average function $\tilde{f}_H (r,s)$.  The generalized variance $\sigma_{DMA}(s)$ is estimated over sub-cubes with size $n_1 \times n_2  \times n_3$ and, then, summed over the whole domain $N_1 \times N_2 \times N_3$ of the fractal cube. The value of $\sigma_{DMA}(s)$ for each sub-cube  is then plotted in log-log scale as a function of $s=\sqrt{n_1^2 + n_2^2 +n_3^2}$. The linearity of this plot guarantees the power-law dependence of the correlation over the investigated range of scales. The slope of the plot yields the Hurst exponent $H$  of the fractional Brownian field.
\end{enumerate}

\par

\section{Generation of complex heterogeneous media}
\label{sect:cube}

The random midpoint displacement  algorithm is a recursive technique widely used for generating fractal series and surfaces.
In $d=1$, a fractional Brownian walk is obtained starting from a line of length $N$. At each iteration $j$, the value at the midpoint is calculated as the average of the two endpoints plus a correction which scales as the inverse of the length of half segment with exponent $H$. The fractal dimension is $D=2-H$, varying between $1$ and $2$. Fractal surfaces with desired roughness can be generated starting from a plane,  whose domain is a regularly spaced square lattice with size $N_1\times N_2$, $i_1=1,2,...,\,N_1$ and $i_2=1,2,...,\,N_2$. First, the square is divided in four subsquares. Then, the value in the center of the square is calculated as the average of the values at the four vertices plus a random term.  Then, the four values at the midpoint of the edges are obtained as the average of the values at the two adjacent vertices plus a random term. The process is repeated until the fractal surface is obtained. The fractal dimension is $D=3-H$, varying between $2$ and $3$.
\par
Here, a high-dimensional implementation of the  algorithm is presented for obtaining a compact body, exhibiting power-law correlation over a wide span of  scales.
\subsection{d-dimensional random midpoint displacement algorithm }
\noindent
Here, the random midpoint displacement algorithm is generalized to be operated  on arrays with arbitrary euclidean dimension $d$.  The procedure is generalized by defining the function:
\begin{equation}\label{md}
f_{H}({r})=\frac{1}{2^d}\sum_k f_k ({r}) + \sigma_{j,d}  \hspace*{10 pt},
\end{equation}
which is defined at the center of a hypercubic equally spaced lattice. The sum is calculated over the $k$ endpoints of the lattice. The quantity $\sigma_{j,d}$ is a random variable defined at each iteration $j$ whose explicit expression is worked out below. We start from Eq.~(\ref{variance}), that is  written as:
\begin{equation}
\left \langle [f_H({{r}}+ {\lambda})-f_H({r})]^2\right \rangle \,\,= \sigma_o^2\left(\sqrt{\lambda_1^2+\lambda_2^2+...+\lambda_d^2}\right)
^{2H} \hspace*{10 pt}, \label{rmd11}
\end{equation}
by considering a hypercubic lattice with size $N_1=N_2=...=N_d=N$, Eq.~(\ref{rmd11}) writes as:
\begin{equation}
\left \langle [f_H({{r}}+ {N})-f_H({r})]^2 \right \rangle \,\, = \sigma_o^2 \left ( \sqrt{d} N \right)
^{2H} \hspace*{10 pt}. \label{rmd22}
\end{equation}
By dividing each lattice size by a factor 2 at each iteration, Eq.~(\ref{rmd22}) is rewritten:
\begin{equation}
\left \langle [f_H({{r}}+ \frac{{N}}{2^j})-f_H({r})]^2 \right \rangle \,\, =  \sigma_o^2 \left( \frac{ \sqrt{d} N }{2^j}\right)^{2H} \hspace*{10 pt}. \label{rmd33}
\end{equation}
Moreover,  at each iteration, the following relation holds:
\begin{equation}
\left \langle [f_H({{r}}+ \frac{{N}}{2^j})-f_H({r})]^2 \right \rangle = \frac{1}{4^{d}}\left \langle [f_H({{r}}+ \frac{{N}}{2^{j-1}})-f_H({r})]^2 \right \rangle  + \sigma_{j,d}^2    \hspace*{10 pt}. \label{rmd44}
\end{equation}
because the value of $f_H(r)$ at each step of the RMD algorithm is calculated over two points for $d=1$ (extremes of a segment), four points for $d=2$ (vertices of a square), eight points for $d=3$  and so on.
By using Eq.~(\ref{rmd33}), Eq.~(\ref{rmd44}) becomes:
\begin{equation}
 \sigma_o^2\left(\frac{ \sqrt{d}N}{2^j}\right)^{2H}  =  \frac{ \sigma_o^2}{4^{d}}\left( \frac{ \sqrt{d} N}{2^{j-1}}\right)^{2H}   + \sigma_{j,d}^2    \hspace*{10 pt}, \label{rmd5}
\end{equation}
and the  term $\sigma_{j,d}^2$  writes:
\begin{equation}
\sigma_{j,d}^2  =  \sigma_o^2 \left( \frac{ \sqrt{d}N}{2^j}\right)^{2H}   \left[ 1-{2^{2(H-d)}}\right]     \hspace*{10 pt}, \label{rmd6}
\end{equation}
which is the generalized form of the relationship holding for $d=1$ \cite{Voss}.

\subsection{Three-dimensional media}
Here, Eqs. (\ref{md},\ref{rmd6}) are used to yield a compact random fractal with embedded Euclidean dimension $d=3$. As already stated, the relevant property of the disordered medium is defined by the scalar function $f_H({r}):\mathbb{R}^3\rightarrow \mathbb{R}$, with ${r}=(x_1, x_2, x_3)$. To generate the fractal structure, a cubic array with size $N_1 \times N_2\times N_3$ is defined. Initially, the cube is fully homogeneous with the function describing the fractal property of the medium is taken as a constant, e.g. $f_H({r})=0$.

\begin{figure}
\centering
\subfigure[\label{a}]%
{\includegraphics[width=3cm,height=2.5cm,angle=0]{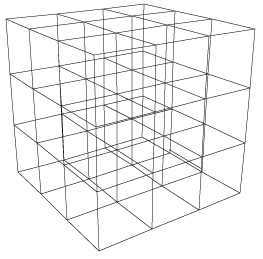}}
\subfigure[\label{b}]%
{\includegraphics[width=3cm,height=2.5cm,angle=0]{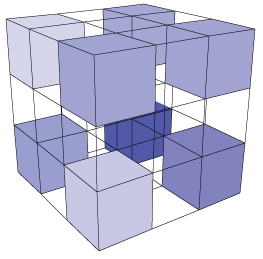}}
\subfigure[\label{c}]%
{\includegraphics[width=3cm,height=2.5cm,angle=0]{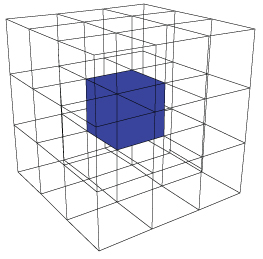}}
\subfigure[\label{d}]%
{\includegraphics[width=3cm,height=2.5cm,angle=0]{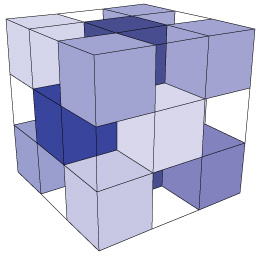}}
\subfigure[\label{e}]%
{\includegraphics[width=3cm,height=2.5cm,angle=0]{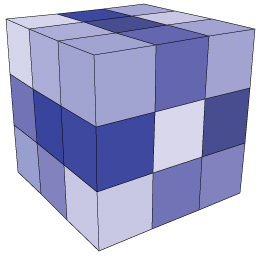}}
\caption{\label{Figure1}
(Color online) Scheme of the elementary steps of the implementation described in Section II.  Step 1: the homogeneous cube is divided in 27 subcubes (a). Step 2: the function is calculated at the subcubes located  at the eight  vertices (b). Step 3: the function is calculated at the subcube located in the center (c).  Step 4: the function is calculated  at subcubes located at the center of the six  faces (d). Step 5: the function is calculated at the  subcubes located at the midpoint of the 12 edges.}
\end{figure}
Then, the algorithm is implemented at each iteration $j$ according to the following  steps:

\begin{description}
\item[Step 1.] The cube is divided in $27$ subcubes, as shown in Fig.~\ref{Figure1}(a).

\item[Step 2.] The  values of the function $f_{H,j}({r})$  are seeded as  random variables of average value $0$ and standard deviation $\sigma_{j,3}$ at the eight subcubes  located at the eight vertices, as shown in Fig.~\ref{Figure1}(b).

 \item[Step 3.] The values of the function $f_{H,j}({r})$, at the central sub-cube, are calculated as follows:
\begin{equation}\label{md3}
f_{H,j}({r})=\frac{1}{8}\sum_{k=1}^8 f_k ({r}) + \sigma_{j,3}  \hspace*{10 pt},
\end{equation}
where the sum is performed over the $k$-values taken by the function $f({r})$ at the subcubes located at the eight vertices of the main cube. The random variable $\sigma_{j,3}$ is defined as:
\begin{equation}
\sigma_{j,3}^2  =  \sigma_o^2 \left( \frac{\sqrt{3}N}{2^j}\right)^{2H}   \left[ 1-{2^{2(H-3)}}\right]     \hspace*{10 pt}. \label{md33}
\end{equation}
 \item[Step 4.] The values of the function $f_{H,j}({r})$, defined over the subcubes located at the center of the six faces, are calculated as follows:%
\begin{equation}\label{md2}
f_{H,j}({r})=\frac{1}{4}\sum_{k=1}^4 f_k ({r}) + \sigma_{j,2}  \hspace*{10 pt},
\end{equation}
where the sum is performed over the $k$-values taken by the function $f_{H,j}({r})$ at the subcubes placed at the four vertices of each face.  The random variable $\sigma_{j,2}$ is defined as:
\begin{equation}
\sigma_{j,2}^2  =  \sigma_o^2 \left( \frac{ \sqrt{2}N}{2^j}\right)^{2H}   \left[ 1-{2^{2(H-2)}}\right]     \hspace*{10 pt}. \label{rmd2}
\end{equation}
%
%

 \item[Step 5.] The  values of the function $f_{H,j}({r})$  at the subcubes located at the midpoint of the twelve edges, are calculated according to:
\begin{equation}\label{md1}
f_{H,j}(r)=\frac{1}{2}\sum_{k=1}^2 f_k ({r}) +\sigma_{j,1} \hspace*{10 pt},
\end{equation}

where the sum is performed over the $k$-values taken by the function $f({r})$ at the subcubes located at the endpoints of the 12 edges.  The random variable $\sigma_{j,1}$ is defined as:
\begin{equation}
\sigma_{j,1}^2  =  \sigma_o^2 \left( \frac{N}{2^j}\right)^{2H}   \left[ 1-{2^{2(H-1)}}\right]     \hspace*{10 pt}.  \label{rmd1}
\end{equation}
Hence, the function $f_{H,j}({r})$ at the subcubes located at the midpoint of each edge takes  a value given by the average of the function  at the endpoint subcubes plus  the random variable $\sigma_{j,1}$.
\end{description}

\begin{figure}[htbp]
\centering
\subfigure[\label{a1}]%
{\includegraphics[width=4.6cm,angle=0]{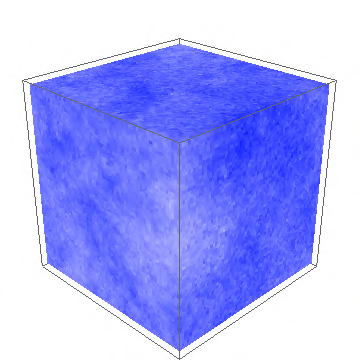}}\hspace{2.27em}%
\subfigure[\label{b1}]%
{\includegraphics[width=4.6cm,angle=0]{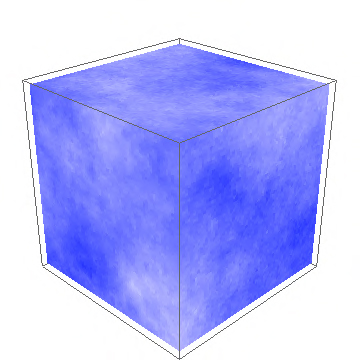}}\hspace{2.27em}%
\subfigure[\label{c1}]%
{\includegraphics[width=4.6cm,angle=0]{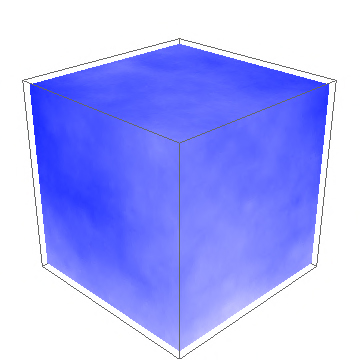}}\hspace{2.27em}%
\caption{\label{Figure2} (Color online) Fractal media generated according to the procedure reported in Section II.  The Hurst exponents are respectively $H=0.2$ (a), $H=0.5$ (b), $H=0.8$ (c). The color map is rescaled in order that white dots correspond to the minimum values and darkest blue dots to the maximum values of $f_H({r})$.}
\end{figure}

The first run of the routine results in 27 subcubes, characterized by the values of the function $f_H(r)$ described above. The structure obtained at the first iteration is shown in Fig.~\ref{Figure1}(e). The  steps 1-5 are iteratively repeated for each of the $27$ subcubes. At the second run, a number of subcubes equal to $27 \times 27$ is obtained. Eventually, the number of subcubes will be  equal to $(3^j)^d$, where $j$ is the iteration number and $d=3$.
\par In Fig.~\ref{Figure2}, the fractal cubes with $H=0.2$ (a), $H=0.5$ (b) and $H=0.8$ (c) are shown at iteration $j=9$. The Hurst exponent $H$ is the input of the generator, which determines the heterogeneity of the resulting microstructure. The color map is rescaled in such a way that the lightest cubes corresponds to the value of  $f_H({r})$ at the initial stage of the medium, darker dots to the maximum values obtained by applying the routine to the function $f_H({r})$.
 \par
\begin{figure}
\centering
\subfigure[\label{a2}]%
{\includegraphics[width=5.5cm,angle=0]{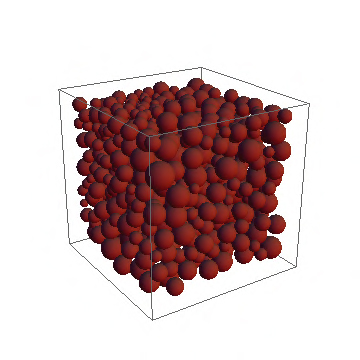}}
\subfigure[\label{b2}]%
{\includegraphics[width=5.5cm,angle=0]{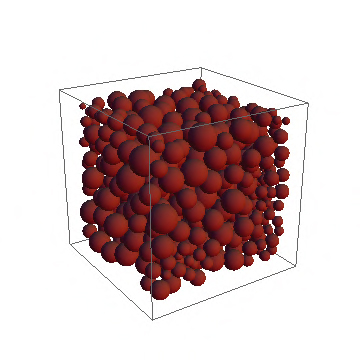}}
\subfigure[\label{c2}]%
{\includegraphics[width=5.5cm,angle=0]{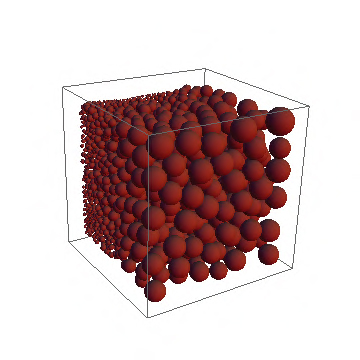}}
\caption{\label{Figure3} (Color online) Granular structures obtained by considering spheres with the radii described by the function $f_H(r)$. The Hurst exponents are respectively $H=0.2$ (a), $H=0.5$ (b) and $H=0.8$ (c). }
\end{figure}
\begin{figure}
\centering
\subfigure[\label{a3}]%
{\includegraphics[width=4.5cm,angle=0]{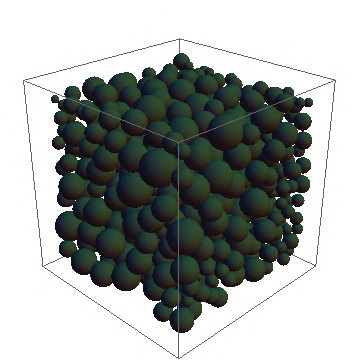}}\hspace{2.7em}%
\subfigure[\label{b3}]%
{\includegraphics[width=4.5cm,angle=0]{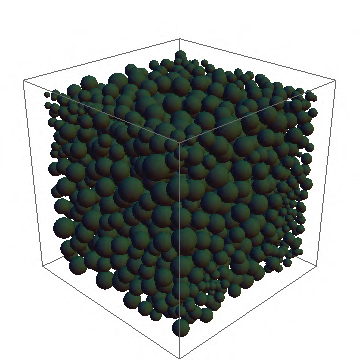}}\hspace{2.7em}%
\subfigure[\label{c3}]%
{\includegraphics[width=4.5cm,angle=0]{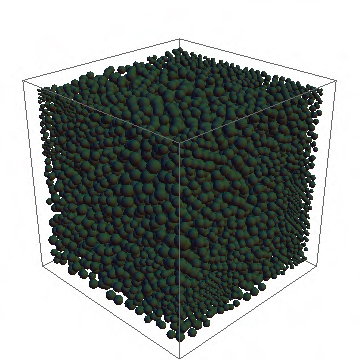}}\hspace{2.7em}%
\caption{\label{Figure4} (Color online) Granular structures obtained by considering spheres with the radii described by the function $f_H(r)$ with three different average grain size. The Hurst exponent is  $H=0.5$ for all the three cases. }
\end{figure}

 An alternative representation can be obtained by relating the fractional Brownian function $f_H({r})$ to the grain size. Granular structures with different Hurst exponents are shown in Fig.~\ref{Figure3}. In this case, since the fractal property is referred to the size of the grain, whose shape is spherical. One can notice that small and large grains are located together for anticorrelated cubes with $H=0.2$. Conversely, as the Hurst exponent increases, the grains segregate according to the size. Smaller grains are more likely to be grouped with small grains and viceversa. In Fig.~\ref{Figure4}, granular structures with the same Hurst exponent (H=0.5) and different average grain size are also shown.

\section{Estimating the Hurst exponent of Random Fractal Media}
\label{sect:dma}
The aim of this section is to implement an independent measure of the  Hurst exponent of the heterogeneous structure generated according to the procedure in Section II (shown in Figs.\,\ref{Figure2},\,\ref{Figure3},\,\ref{Figure4}.  The detrending moving average algorithm has been proposed in \cite{Carbone1} for estimating the Hurst exponent of high-dimensional ($d\geq 2$) fractals. In the present work, the algorithm will be implemented on the heterogeneous structures generated in Section II.
\par
The core of the algorithm is the generalized variance  $\sigma^2_{DMA}(s)$, that for a three-dimensional structure, is written:
\begin{equation}
 \sigma^2_{DMA}(s)=\frac{1}{{V}}\sum_{V}\Big[f_H(r)-\widetilde{f}_{n_1,n_2,n_3}(r)\Big]^2  \hspace*{3 pt},\label{DMAd}
\end{equation}
\noindent  where $f_H(r)=f_H(x_1,x_2,x_3)$  is the fractional
Brownian field with $i_1=1,2,...,\,N$, $i_2=1,2,...,\,N$,
$i_3=1,2,...,\,N$.
The function $\widetilde{f}_{n_1,n_2,n_3}(r)$ is given by:

\begin{eqnarray}
\label{MAd} \widetilde
f_{n_1,\,n_2,\,n_3}(r)=\frac{1}{\nu}  \sum_{k_1} \sum_{k_2}\sum_{k_3} f_H(x_1-k_1,x_2-k_2,x_3-k_3)
\hspace*{3 pt} ,
\end{eqnarray}
with  the size of the  subcubes $(n_1,\,n_2,\,n_3)$  ranging from $(3,3,3)$ to the  maximum values $(n_{1max},\,n_{2max},\,n_{3max})$.   $\nu=n_1 n_2 n_3$ is the volume of the subcubes.
The quantity ${{V}}={(N_1-n_{1max})(N_2-n_{2max})(N_3-n_{3max})}$ is the volume of the fractal cube over which the averages $\tilde{f}$ are defined.
As observed in \cite{Carbone1},  Eqs.~(\ref{DMAd}) and (\ref{MAd}) are defined for any   geometry of the
sub-arrays. However, in practice, sub-cubes  with $n_1=n_2=n_3$ are computationally more suitable for avoiding spurious directionality and biases in the calculations. The generalized variance $\sigma^2_{DMA}(s)$  scales as
$(n_1^2+n_2^2+n_3^2\,)^{H}$  because of the fundamental
property  of fractional Brownian functions Eq.~(\ref{variance}).
\par
Eqs.~(\ref{DMAd}) and (\ref{MAd}) correspond to the isotropic implementation of the algorithm. The isotropy follows from the definition of the average $\widetilde{f}_{n_1,n_2,n_3}(r)$, which is obtained by summing all the values taken by $f_H(r)$ at the subcubes centered in $r$. As explained in \cite{Carbone1}, the implementation can  be made anisotropic for fractals having a preferential growth direction, as for example biological tissues, epitaxial layers, crack propagation.
The anisotropy is accomplished by varying the sum indexes according to
$m_1=\mathrm{int}(n_1
\theta_1)$,\,$m_2=\mathrm{int}(n_2
\theta_2)$,\, $m_3=\mathrm{int}(n_3 \theta_3)$.
Upon variation of the parameters $\theta_1$,
$\theta_2$,  $\theta_3$ in the range $\left[ 0, 1 \right]$
the indexes  of
the sums in  Eqs.~(\ref{DMAd}) and (\ref{MAd}) are
set within  each subcube. In particular,  $r$
coincides respectively with (a) one of the vertices
for $\theta_1,\theta_2,\theta_3=$ $0$ or
$1$ or with (b) the center
for $\theta_1=\theta_2=\theta_3=1/2$. The values $\theta_1=\theta_2=\theta_3=1/2$  correspond
to the \emph{isotropic} implementation, while
$\theta_1=\theta_2=\theta_3=0$ and
$\theta_1=\theta_2=\theta_3=1$ correspond to the
\emph{directed} implementation. In $d=3$, the
\emph{isotropic} implementation implies that the function defined by  Eq.~(\ref{MAd}) is  calculated over subcubes  whose center is
$r$. Conversely, the \emph{anisotropic} implementation
implies that the function $\widetilde{f}_{n_1,n_2,n_3}(r)$ is calculated over
subcubes  having one of the eight vertices coinciding with $r$\,.
\begin{figure}
\includegraphics[width=12cm,height=9cm,angle=0]{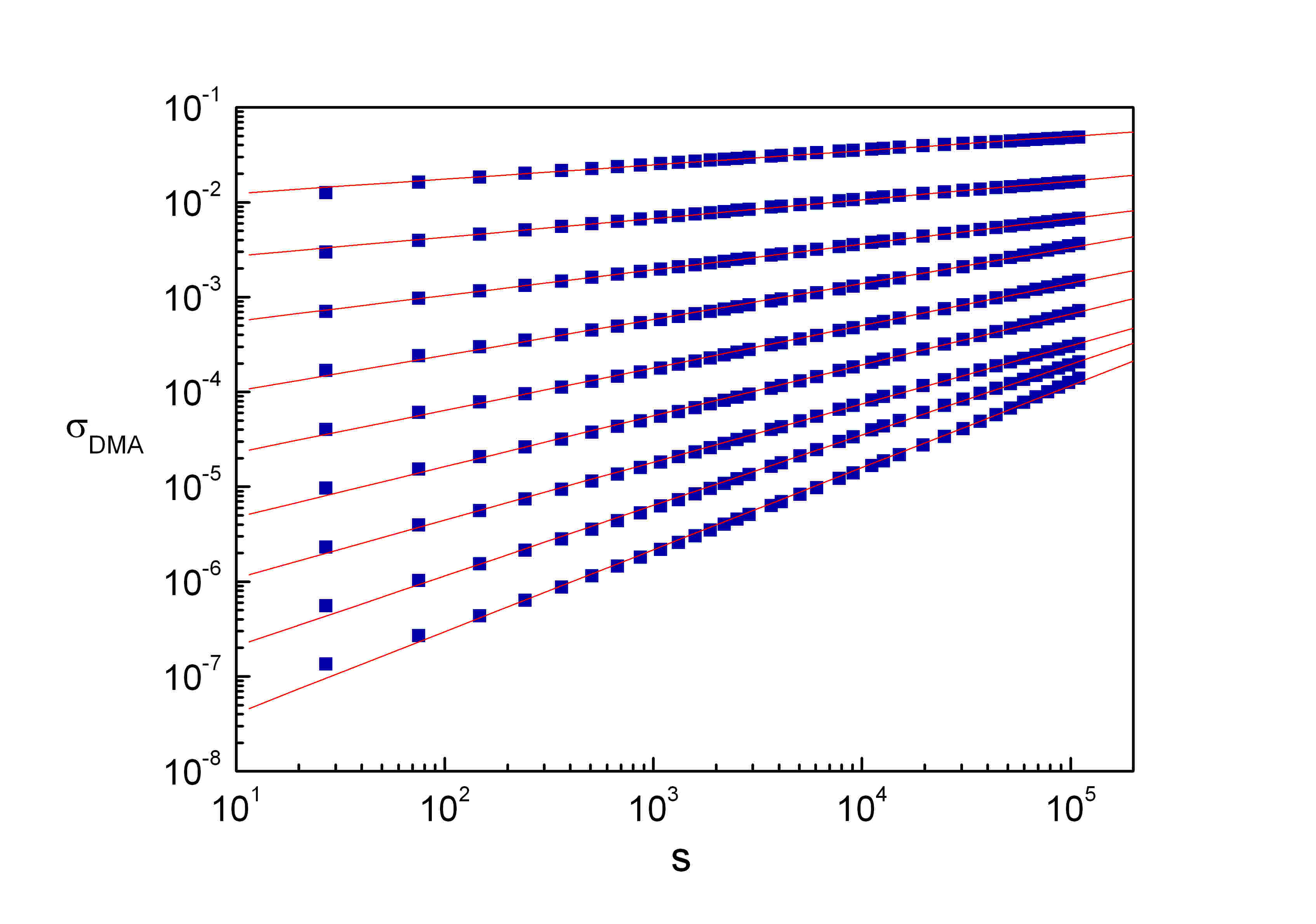}
\caption{\label{Figure5} (Color online)  Log-log plot of
$\sigma^2_{DMA}(s)$ for fractal media with size $N_1 \times N_2 \times N_3=1025 \times
1025 \times 1025$. The fractal media are generated by the algorithm proposed in Section II, with  Hurst exponent varying from $0.1$ to $0.9$
with step $0.1$. Dashed lines represent the linear fits. The Hurst exponents can be estimated by means of the 3d-DMA algorithm as explained in Section III. }
\end{figure}
\begin{table}
\caption{\label{tab:variance} Hurst exponents $H_{1}$, $H_{2}$, $H_{3}$, $H_{4}$ and
linear regression coefficients $\rho_1$, $\rho_2$, $\rho_3$, $\rho_4$
of curves like those of Fig.~\ref{Figure5} obtained as average of 10 realizations of heterogeneous media with $N_1
\times N_2\times N_3=1025 \times 1025 \times 1025$.  $H_{1}$ and $H_{2}$ have been calculated by
linear fit over the full range of $s$. $H_{3}$ and $H_{4}$ have been calculated by
linear fit over the range $10^2\leq s \leq 10^4$. One can observe that the quality of the fit improves in the central range of scales.}
\begin{ruledtabular}
\begin{tabular}{ccccccccc}
  H & H$_1$   & $\rho_1$ & H$_2$   & $\rho_2$ & H$_3$   & $\rho_3$  &  H$_4$   & $\rho_4$    \\
  \hline
0.1  & 0.1503    & 0.9962    & 0.1470  & 0.9965   &  0.1520 & 0.9998  & 0.1445  & 0.9995\\  \hline
0.2  & 0.1973	 & 0.9984	 & 0.1939  & 0.9984   &  0.2005 & 0.9999  & 0.2028  & 0.9999 \\  \hline
0.3  & 0.2700	 & 0.9998    & 0.2688  & 0.9998   &  0.2738 & 0.9999  & 0.2809  &0.9996 \\  \hline
0.4  & 0.3780	 & 0.9984	 & 0.3881  & 0.9980   &  0.3718 & 0.9993  & 0.3558  & 0.9997  \\  \hline
0.5  & 0.4467	 & 0.9994	 & 0.4513  & 0.9994   &  0.4514 &  0.9997 & 0.4656  & 0.9993  \\ \hline
0.6  & 0.5362	 & 0.9992	 & 0.5418  & 0.9993   &  0.5494 & 0.9996  & 0.5426  & 0.9997\\ \hline
0.7  & 0.6129    & 0.9996	 & 0.6115  & 0.9997   &  0.6349 & 0.9998  & 0.6877  & 0.9992 \\  \hline
0.8  & 0.7430	 & 0.9993	 & 0.7412  & 0.9995   &  0.7741 & 0.9997  & 0.7863  & 0.9997   \\  \hline
0.9  & 0.8645	 & 0.9990	 & 0.8747  & 0.9991   &  0.8650 & 0.9999  & 0.8775  & 0.9999  \\
\end{tabular}
\end{ruledtabular}
\end{table}

\par In order to calculate the Hurst exponent of the heterogeneous structure, the algorithm
is implemented through the following steps. The function
$\widetilde{f}_{n_1,n_2,n_3}(r)$ is calculated over different subcubes,
by varying $n_1,\, n_2,\, n_3$ from $3\times3\times3$ to $n_{1max}\times n_{2max}\times n_{3max}$. The maximum values
$n_{1max}\,,n_{2max}\,,n_{3max}$ depend on the  size of
the whole fractal. To minimize finite-size effects, it should be $n_{1max}<<N_1$,
$n_{2max}<<N_2$, $\,n_{3max}<<N_3$.
\noindent
The next step  is the calculation of the difference $f_H(r)-\widetilde
f_{n_1,n_2,n_3}(r)$ in parentheses of Eq.~(\ref{DMAd}) for each sub-cube $n_1\times n_2 \times n_3$.
 For each
subcube, the corresponding value of $\sigma_{DMA}^2 (s)$
is calculated and finally plotted on log-log axes.  The
log-log plot of $\sigma_{DMA}^2 (s)$ as a function of  $s$\,, yields a straight
line with slope $H$, on account of the following relationship:
\begin{equation}
\sigma_{DMA}^{2}(s) \sim \left(n_1^2+n_2^2+n_3^2 \right)^{H}\sim {s}^{H} \hspace*{5 pt} .
\label{sigma2b}
\end{equation}

\noindent
In Fig.~\ref{Figure5}, the log-log plots of $\sigma_{DMA}^2 (s)$ vs. $s$ are shown for fractal cubes
generated  according to the procedure described in  Section II.
The cubes have
Hurst exponents  ranging from $0.1$ to $0.9$ with step
$0.1$ and size  $1025 \times 1025 \times 1025$.
 Dashed lines
represent the linear fits, with errors and linear regression coefficients shown in Table I.  The plots
of $\sigma^2_{DMA}(s)$ as a function  $s$ are linear according to
the power-law behavior expected on the basis of Eq.~(\ref{sigma2b}).
Deviations from the full linearity
can be observed particularly  at the extremes
of the scale.
\section{Conclusions}
We have put forward an algorithm to generate a fully compact heterogeneous medium,  whose fractal dimension can be continuously varied  upon varying the Hurst exponent between $0$ and $1$. The generation method is based on a generalization of the random midpoint displacement algorithm. In order to check the accuracy of the generator, the Hurst exponent of the fractals can be estimated by using
the high-dimensional
variance $\sigma^2_{DMA}(s)$ defined by
Eq.~(\ref{DMAd}). We envisage fruitful applications, e.g. in three-dimensional medical image analysis, where density or granularity patterns could be interpreted synthetically by means of fractal descriptors.
Generally,  such a  structure can properly reproduce complex systems whose heterogeneity is described by correlation decaying as a power law over space.  This correlation corresponds to the intermediate behavior of real structure, as opposed to the fully uncorrelated exponential decay and the fully correlated constant decay exhibited respectively by the correlation function of ideal cases such as the perfect gas and the regular lattice.

\par

\section{acknowledgements}
We acknowledge financial support by Regione Piemonte and Politecnico di Torino.
CINECA is gratefully acknowledged for CPU time at the High Performance Computing (HPC) environment.


\begin{thebibliography}{<50>}
\bibitem{nota}
The Menger sponge is  obtained by dividing a cube into $27$ identical cubic pieces. Then, seven cubic pieces are removed from the center of the cube and from the center of each of the six faces. The procedure is then repeated to the $20$ remaining cubes.
Ideally, the procedure should be repeated infinite times.   The Menger sponge has fractal dimension  $D=\log 20/\log 3=2.7268$ and volume decreasing at each iteration as $V_l=(20/27)^l$. The Menger sponge, at infinite iterations, is a non compact three dimensional fractal structure having infinite area and zero-volume.
\end{thebibliography}
\end{document}